\documentclass[12pt]{article}   
\usepackage{epsfig}

\newlength{\dinwidth} \newlength{\dinmargin}
\begin{document}
\begin {flushright}
Cavendish-HEP-04/13\\
DESY-04-071
\end {flushright} 
\vspace{3mm}

\begin{center}
{\Large \bf $W$ Boson Production at Large Transverse 
Momentum\footnote{Presented
at the DIS 2004 Workshop, Strbske Pleso, Slovakia, 14-18 April, 2004.}}
\end{center}
\vspace{2mm}

\begin{center}
{\large Nikolaos Kidonakis$^a$ and 
\underline{Agust{\' \i}n Sabio Vera$^b$}}\\
\vspace{2mm}
{\it $^a$ Cavendish Laboratory, University of Cambridge\\
Madingley Road, Cambridge CB3 0HE, UK\\
E-mail: kidonaki@hep.phy.cam.ac.uk\\
\vspace{1mm}
$^b$ II. Institut f{\"u}r Theoretische Physik, Universit{\"a}t Hamburg\\ 
Luruper Chaussee 149, 22761~Hamburg, Germany\\
E-mail: sabio@hep.phy.cam.ac.uk}
\end{center}

\vspace{3mm}

\begin{abstract}

The production of $W$ bosons at large transverse momentum
in $p{\bar p}$ collisions is studied. The next-to-leading order cross 
section in this region is dominated by threshold soft-gluon corrections.
The transverse momentum distribution of the $W$ at the Tevatron is 
modestly enhanced when next-to-next-to-leading-order soft-gluon corrections
are added, and the dependence on the factorization 
and renormalization scales is significantly reduced.

\end{abstract}

\thispagestyle{empty} \newpage \setcounter{page}{2}

\section{Introduction}

$W$ boson hadroproduction is a process of importance in testing the 
Standard Model and in estimates of backgrounds to new physics, such 
as associated Higgs boson production at the Tevatron. 
In Refs.~\cite{AR,gpw} the complete next-to-leading-order (NLO)
cross section for $W$ hadroproduction at large transverse momentum was 
first calculated. The NLO corrections contribute an enhancement of the 
differential distributions in transverse momentum $Q_T$ of the $W$ boson 
and considerably stabilize the dependence of the cross section on the 
factorization and renormalization scales.

$W$ boson production at the Tevatron receives important contributions
from the near-threshold kinematical region.
The calculation of hard-scattering cross sections near partonic threshold 
involves corrections from the emission of soft gluons from the partons in 
the process. At each order in perturbation theory there are large 
logarithms arising from incomplete cancellations near partonic
threshold between graphs with real emission and virtual graphs due to the 
limited phase space available for real gluon emission. 
These threshold corrections exponentiate as a result of the factorization 
properties~\cite{KS,KOS,LOS,NKtop} of the cross section and have 
been successfully resummed for a large number of processes~\cite{NK}.

In Ref.~\cite{NKVD} the resummation of threshold logarithms
for $W$ boson hadroproduction at large transverse momentum  
was first studied, and the expansion of the resummed cross section at
next-to-next-to-leading order (NNLO) and 
next-to-next-to-leading logarithmic (NNLL) accuracy was presented.  
Following Ref.~\cite{NKuni}, the accuracy of the theoretical prediction was 
increased in Ref.~\cite{Kidonakis:2003xm} to include 
next-to-next-to-next-to-leading logarithms (NNNLL) and
phenomenological studies
for $W$ production at the Tevatron were also presented. 
We note that the theoretical and numerical results are similar to those found 
for direct photon production~\cite{NKJO}. 

\section{Theoretical results}

For the process $ h_A(P_A)+h_B(P_B) \longrightarrow W(Q) + X ,$ 
we can write:
\begin{eqnarray}
E_Q\,\frac{d\sigma_{h_Ah_B{\rightarrow}W(Q)+X}}{d^3Q} &=&
\sum_{f} \int dx_1 \, dx_2 \; \phi_{f_a/h_A}(x_1,\mu_F^2) 
\; \phi_{f_b/h_B}(x_2,\mu_F^2) 
\nonumber \\ && \hspace{-10mm} \times \,
E_Q\,\frac{d\hat{\sigma}_{f_af_b{\rightarrow}W(Q)+X}}{d^3Q}
(s,t,u,Q,\mu_F,\alpha_s(\mu_R^2)) \, , \nonumber
\end{eqnarray}
where $E_Q=Q^0$, $\phi_{f/h}$ are the parton distributions, 
$\hat{\sigma}$ is the parton-level cross section, $\mu_F$ is  
the factorization scale, $\mu_R$ is the renormalization scale,
and $s,t,u$ are parton-level kinematical invariants. 
The lowest-order parton level subprocesses are 
$q(p_a)+g(p_b) \longrightarrow W(Q) + q(p_c)$ and 
$q(p_a)+{\bar q}(p_b) \longrightarrow W(Q) + g(p_c)$. 
The variable $s_2=(p_a+p_b-Q)^2$ 
is defined as the invariant mass of the system recoiling 
against the $W$ at the parton level. It parametrizes the inelasticity of 
the parton scattering, being zero for one--parton production.

For the $n$-th order corrections in the strong coupling $\alpha_s$,
the partonic cross section $\hat{\sigma}$ 
includes distributions with respect to $s_2$ of the type
$ \left[\ln^{m}(s_2/Q_T^2)/s_2 \right]_+$, with $m\le 2n-1$.
Leading logarithms (LL) are those with $m=2n-1$, next--to--leading 
logarithms (NLL) with $m=2n-2$, 
next--to--next--to--leading logarithms (NNLL) with $m=2n-3$,
and next--to--next--to--next--to--leading logarithms (NNNLL) with $m=2n-4$.

The NLO soft and virtual, and the NNLO soft--gluon 
corrections are presented in the ${\overline{\rm MS}}$ scheme 
as given in Ref.~\cite{Kidonakis:2003xm}.
Here we show the formulae for the $qg \longrightarrow Wq$ subprocess. 
The $q{\bar q} \longrightarrow Wg$ subprocess is calculated in a similar 
way~\cite{Kidonakis:2003xm}. 

The NLO soft and virtual corrections for $qg \rightarrow Wq$ are
\begin{eqnarray}
E_Q\frac{d{\hat\sigma}^{(1)}_{qg \rightarrow Wq}}{d^3Q} = 
F^B_{qg \rightarrow Wq} 
{\alpha_s(\mu_R^2)\over\pi}\,
\left\{c_3^{qg} \, \left[\frac{\ln(s_2/Q_T^2)}{s_2}\right]_+ 
+c_2^{qg} \, \left[\frac{1}{s_2}\right]_+ + c_1^{qg} \, 
\delta(s_2)\right\} , \nonumber
\end{eqnarray}
where $F^B_{qg \rightarrow Wq}$ is the Born term.
The LL $[\ln(s_2/Q_T^2)/s_2]_+$ term 
and the NLL $[1/s_2]_+$ term are the soft gluon corrections.
The $\delta(s_2)$ term gives the virtual corrections.
In the discussion below ``NLO-NLL'' denotes when,
at NLO, all the LL and NLL soft--gluon terms, and also the 
scale-dependent terms in $\delta(s_2)$, are included. 
The NLO coefficients, $c_1^{qg}, c_2^{qg}$ and $c_3^{qg}$, 
can be found in Ref.~\cite{Kidonakis:2003xm}. 

Using the conventions in Ref.~\cite{Kidonakis:2003xm}, 
the NNLO soft and virtual corrections are 
\begin{eqnarray}
E_Q\frac{d{\hat\sigma}^{(2)}_{qg \rightarrow Wq}}{d^3Q} = 
F^B_{qg \rightarrow Wq} 
\frac{\alpha_s^2(\mu_R^2)}{\pi^2} \, 
{\hat{\sigma'}}^{(2)}_{qg \rightarrow Wq} \nonumber
\end{eqnarray}
with
\begin{eqnarray}
{\hat{\sigma'}}^{(2)}_{qg \rightarrow Wq}&=& 
\frac{1}{2} (c_3^{qg})^2 \, \left[\frac{\ln^3(s_2/Q_T^2)}{s_2}\right]_+ 
+\left[\frac{3}{2} c_3^{qg} \, c_2^{qg} 
- \frac{\beta_0}{4} c_3^{qg}
+C_F \frac{\beta_0}{8}\right] \left[\frac{\ln^2(s_2/Q_T^2)}{s_2}\right]_+
\nonumber \\ && \hspace{-15mm}
{}+\left\{c_3^{qg} \, c_1^{qg} +(c_2^{qg})^2
-\zeta_2 \, (c_3^{qg})^2 -\frac{\beta_0}{2} \, T_2^{qg} 
+\frac{\beta_0}{4} c_3^{qg}  \ln\left(\frac{\mu_R^2}{s}\right)
+(C_F+C_A) \, K \right.
\nonumber \\ && \hspace{-15mm} \quad \quad \left.
{}+C_F \left[-\frac{K}{2} 
+\frac{\beta_0}{4} \, \ln\left(\frac{Q_T^2}{s}\right)\right]
-\frac{3}{16} \beta_0 C_F \right\}
\left[\frac{\ln(s_2/Q_T^2)}{s_2}\right]_+
\nonumber \\ && \hspace{-15mm} 
{}+\left\{c_2^{qg} \, c_1^{qg} -\zeta_2 \, c_2^{qg} \, c_3^{qg}
+\zeta_3 \, (c_3^{qg})^2 
-\frac{\beta_0}{2} T_1^{qg}
+\frac{\beta_0}{4}\, c_2^{qg} \ln\left(\frac{\mu_R^2}{s}\right) 
+{\cal G}_{qg}^{(2)}
\right. 
\nonumber \\ && \hspace{-20mm} 
{}+(C_F+C_A) \left[\frac{\beta_0}{8} 
\ln^2\left(\frac{\mu_F^2}{s}\right)
-\frac{K}{2}\ln\left(\frac{\mu_F^2}{s}\right)\right]
-C_F\, K \, \ln\left(\frac{-u}{Q_T^2}\right)
-C_A\, K \, \ln\left(\frac{-t}{Q_T^2}\right)
\nonumber \\ && \hspace{-15mm} \quad \quad \left.
{}+C_F \, \left[\frac{\beta_0}{8}
\ln^2\left(\frac{Q_T^2}{s}\right)
-\frac{K}{2}\ln\left(\frac{Q_T^2}{s}\right)\right]
-\frac{3}{16}\beta_0 C_F\, \ln\left(\frac{Q_T^2}{s}\right) \right\}
\left[\frac{1}{s_2}\right]_+  
\nonumber \\ &&  \hspace{-15mm}
{}+ \left\{R_{qg}^{\mu,\,(2)}+ R_{qg}^{(2)} \right\} \delta(s_2) \, , 
\nonumber
\end{eqnarray}
where the definitions of the functions entering this expression 
can be found in Ref.~\cite{Kidonakis:2003xm}. 
The soft corrections are the LL $[\ln^3(s_2/Q_T^2)/s_2]_+$ term, 
the NLL $[\ln^2(s_2/Q_T^2)/s_2]_+$ term, 
the NNLL $[\ln(s_2/Q_T^2)/s_2]_+$ term, and
the NNNLL $[1/s_2]_+$ term. In ${\cal G}^{(2)}_{q g}$, in the NNNLL term, 
two-loop process-dependent contributions are not included, but, from related 
studies (direct photon production \cite{NKJO}, 
top hadroproduction \cite{NKRV}) they are expected to give a small 
contribution. ``NNLO-NNLL'' will indicate inclusion of the LL, NLL, and NNLL 
terms at NNLO. ``NNLO-NNNLL'' includes the NNNLL terms as well. 
The term $R_{qg}^{\mu,\,(2)} \, \delta(s_2)$ denotes the scale-dependent 
virtual corrections which are given explicitly in Ref.~\cite{Kidonakis:2003xm}.
$R_{qg}^{(2)} \delta(s_2)$ are the scale-independent virtual corrections
which are currently unknown.

\section{Numerical results}

\begin{figure}[!thb]
\vspace*{7.0cm}
\begin{center}
\includegraphics{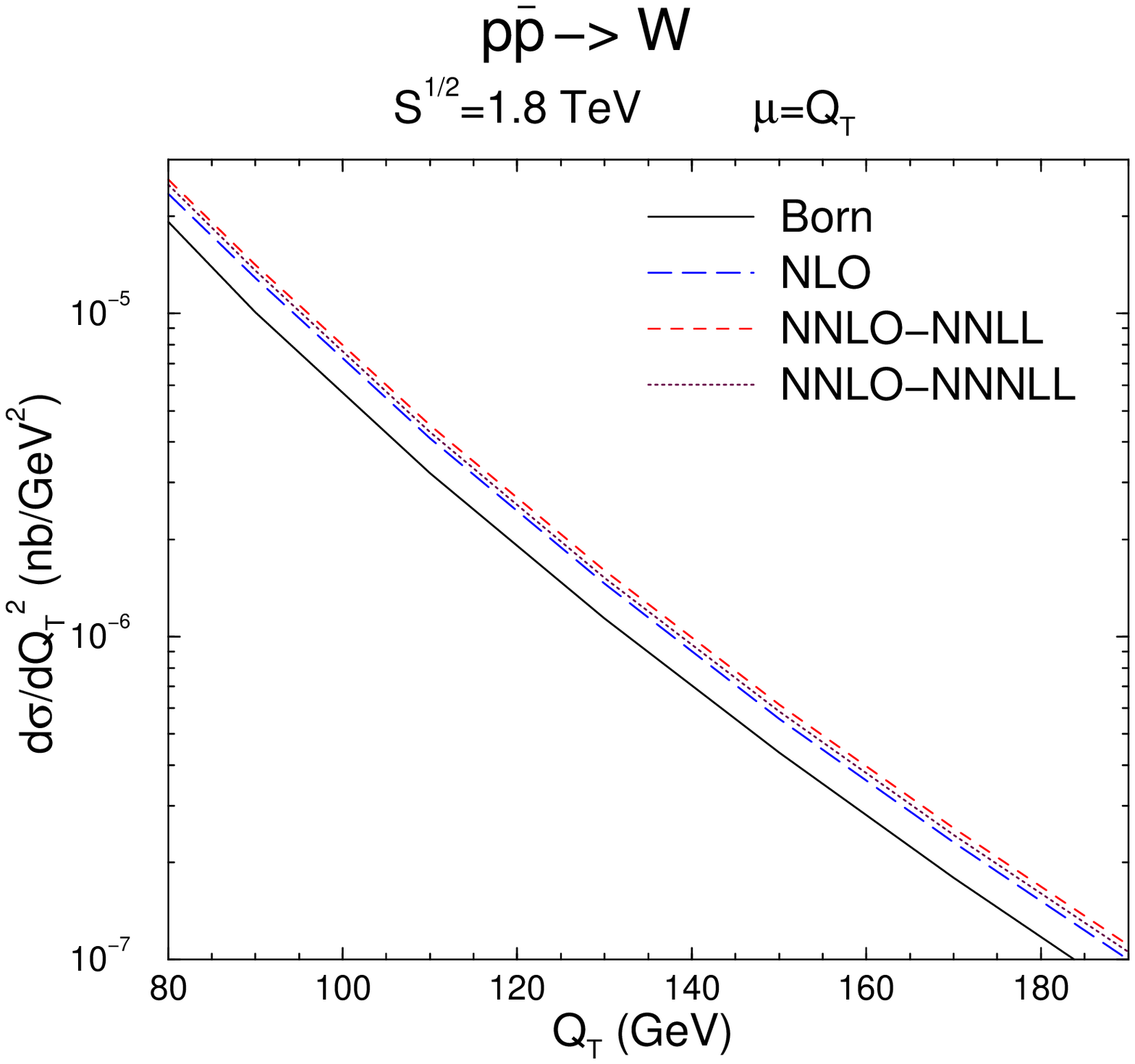}
\includegraphics{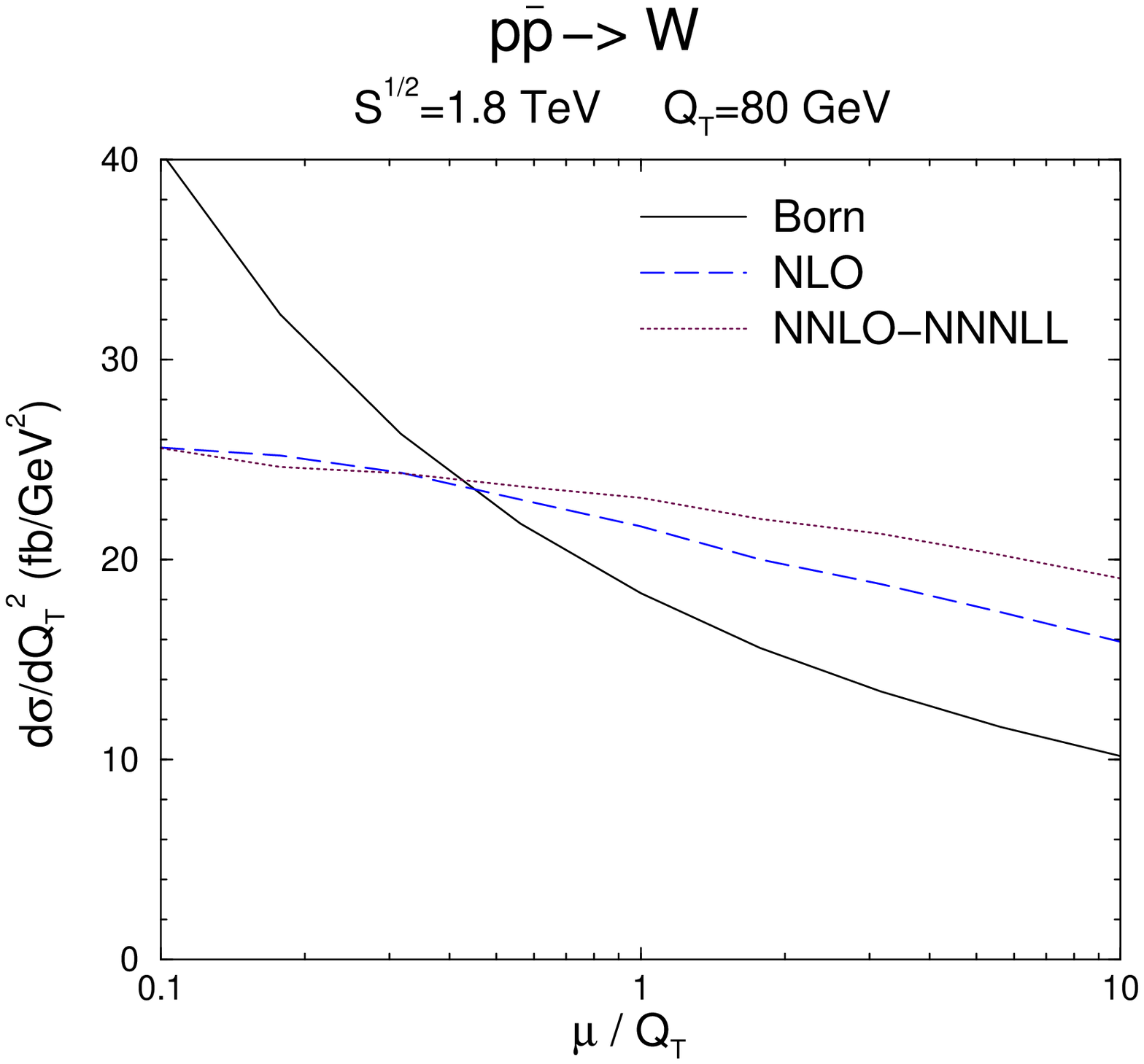}
\caption[*]{$d\sigma/dQ_T^2$ for $W$ production 
at the Tevatron with $\sqrt{S}=1.8$ TeV as a function of
(left) $Q_T$ with $\mu=Q_T$, and (right) $\mu/Q_T$ with $Q_T=80$ GeV.}
\label{fig1}
\end{center}
\end{figure}

The MRST2002 approximate NNLO parton densities \cite{MRST} are used 
in the evaluation of the numerical results.
At the left hand side of Fig.~\ref{fig1} 
we plot the transverse momentum distribution,
$d\sigma/dQ_T^2$, for $W$ hadroproduction at the Tevatron Run I 
with $\sqrt{S}=1.8$ TeV in the high-$Q_T$ region setting $\mu=Q_T$, 
where $\mu \equiv\mu_F=\mu_R$.
We plot Born, exact NLO \cite{AR,gpw}, NNLO--NNLL, and 
NNLO--NNNLL results. The 
NLO corrections provide a significant enhancement of the Born
cross section. The NNLO--NNLL corrections provide a further
modest enhancement of the $Q_T$ distribution. 
The more accurate NNNLL contributions are negative forcing 
the NNLO--NNNLL cross section to lie between
the NLO and NNLO--NNLL results. 

At the right hand side of Fig.~\ref{fig1} the scale dependence of the 
differential cross section for $Q_T=80$ GeV is shown. We plot
the differential cross section versus $\mu/Q_T$ over two
orders of magnitude: $0.1 < \mu/Q_T < 10$. We note the good
stabilization of the cross section when the NLO corrections are
included, and the further improvement when the NNLO-NNNLL corrections
(which include all the soft and virtual NNLO scale terms) are added.
The NNLO-NNNLL result approaches the scale independence expected of a truly
physical cross section. 

\begin{figure}[!thb]
\vspace*{7.0cm}
\begin{center}
\includegraphics{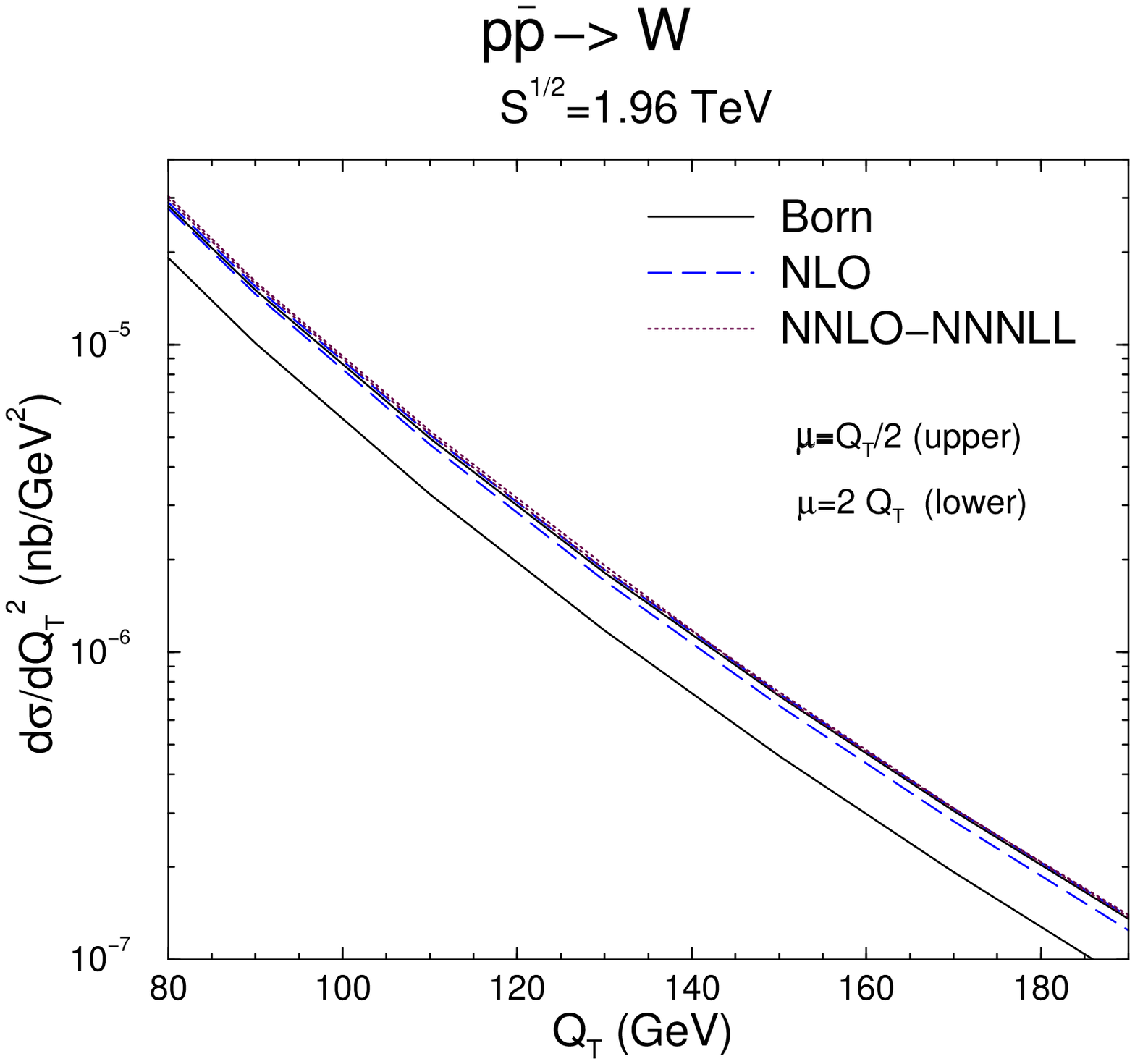}
\includegraphics{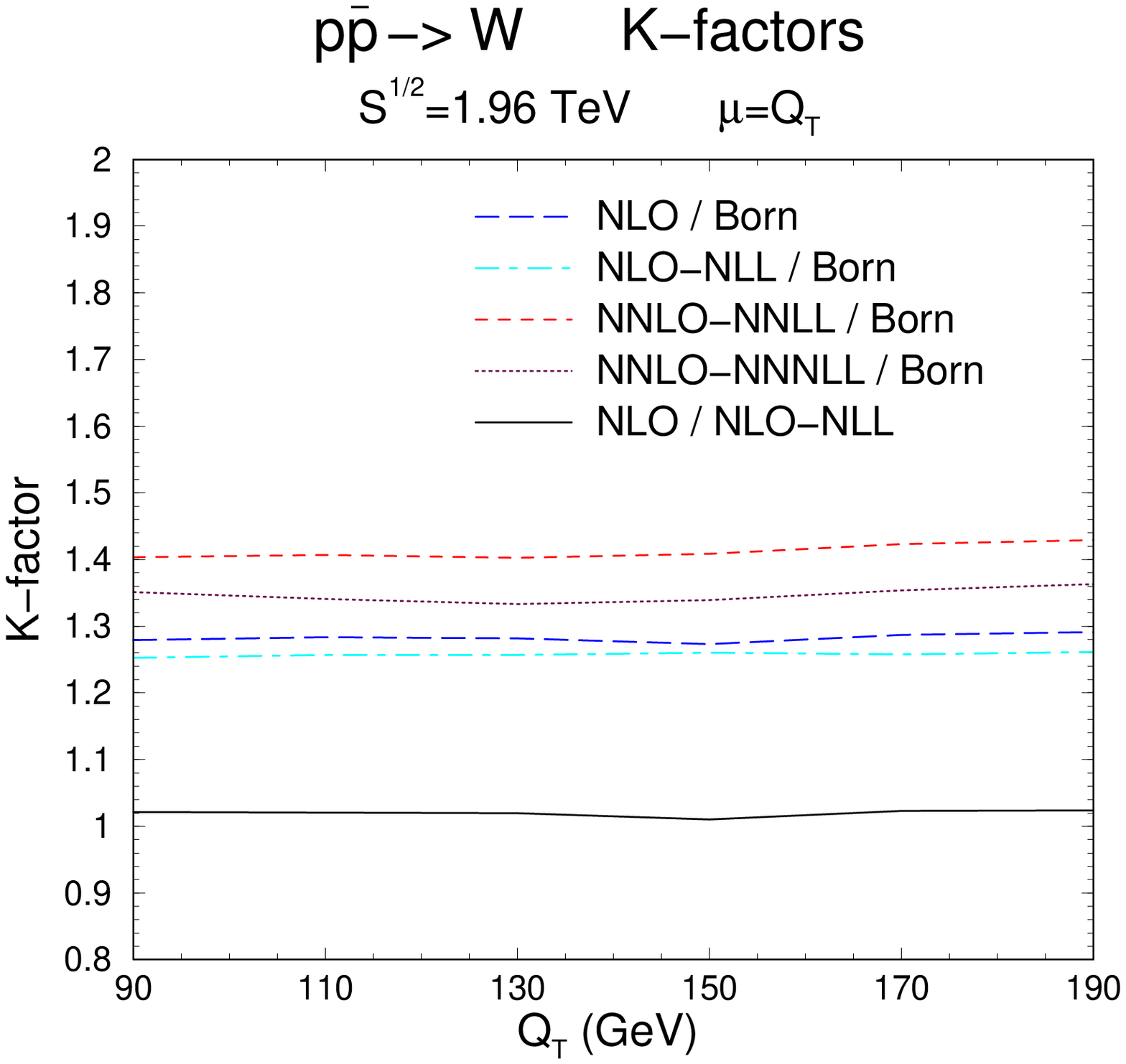}
\caption[*]{Left: $d\sigma/dQ_T^2$ for $W$ production 
at the Tevatron with $\sqrt{S}=1.96$ TeV as a function of 
$Q_T$ with $\mu=Q_T/2, 2Q_T$. Right: the K-factors  
versus $Q_T$ with $\mu=Q_T$.}
\label{fig2}
\end{center}
\end{figure}

In Fig.~\ref{fig2} we show results at the Tevatron Run II 
with $\sqrt{S}=1.96$ TeV. On the left we plot the Born, NLO, and NNLO-NNNLL
cross sections for two choices of scale, $\mu=Q_T/2$ and $2Q_T$.
We see that while the Born result varies a lot between the two scales,
the NLO result is more stable, and the NNLO-NNNLL is so stable that the
two curves are indistinguishable. Note that all $\mu=Q_T/2$ curves
lie on top of each other. This is all as expected from the right
plot in Fig.~\ref{fig1}. 
On the right-hand side of Fig.~\ref{fig2}
are shown the $K$-factors, i.e. the ratios of
cross sections at various orders and accuracies
to the Born cross section, all with $\mu=Q_T$. 
The NLO/NLO-NLL curve also shows that the NLO-NLL result
is a very good approximation to the full NLO result, i.e. the
soft--gluon corrections dominate the NLO cross section. 
The difference between NLO and NLO-NLL is only 2\% for $Q_T > 90$ GeV
and less than 10\% for lower $Q_T$ down to 30 GeV. The fact that the
soft--gluon corrections dominate the NLO cross section
is a major justification for studying the NNLO soft gluon corrections
to this process. 

\section{Conclusions}

The NNLO soft--gluon corrections for $W$ hadroproduction at large transverse 
momentum in $p{\bar p}$ collisions at the Tevatron Run I and II have been 
presented.
The NLO soft--gluon corrections dominate the NLO differential cross section 
in this region,
while the NNLO soft--gluon corrections provide modest enhancements
and further decrease the factorization and renormalization
scale dependence of the transverse momentum distributions.

\section*{Acknowledgements} We thank Richard Gonsalves for help with the 
NLO corrections. The research of N.K. has been 
supported by a Marie Curie Fellowship of the European Community programme 
``Improving Human Research Potential'' under contract no.
HPMF-CT-2001-01221. 
The work of A.S.V. was supported by an Alexander von Humboldt 
Postdoctoral Fellowship.

\end{document}